# On the electrical conductivity of transition metals

T. Rajasekharan and V. Seshubai

[1]Powder Metallurgy Division, Defence Metallurgical Research Laboratory, Hyderabad 500 058, India
[2] School of Physics, University of Hyderabad, Hyderabad 500 046, India

ABSTRACT

We study the consequences of identifying the 'electron density parameter' *N* of Miedema et al. as the accurate valence of metallic elements. The incorporation of these as valence in Pauling's scheme for the electronic structure of transition metals, predicts, for the first time, that the electrical conductivity of transition metals would decrease in the order $\sigma_{Ag} > \sigma_{Cu} > \sigma_{Au} > \sigma_{Rh} > \sigma_{others}$, as is experimentally observed.

___

Corresponding author

E-mail address : trajasekharan@gmail.com

1. **Introduction**

The electrical conductivity of metallic elements has values that span a wide range and is one of the most accurately measured physical quantities. However, the current theoretical approaches, including quantum mechanical calculations using DFT, though accurate in their prediction of cohesive properties, have some difficulties in arriving at the electrical properties of the condensed state; for instance [1, 2], the calculated band gaps in semiconductors can be in error by as much as 100%. As far as we know, there has been no theoretical demonstration till now as to why Ag has the maximum electrical conductivity of all metals, followed by Cu, then Au, then Rh and so on.

Pauling had proposed that [3] in metals, d-orbitals alone are not especially well suited to use in bond formation, but hybridization of d, s, and p orbitals leads to the best bond orbitals known. There are available a total of nine relatively stable orbitals: five 3d, one 4s, and three 4p for the iron group elements, and corresponding sets for other series. According to him, as per the empirical evidence, about 2.44 d orbitals on an average show only weak interatomic interactions, and the remaining 2.56 d orbitals combine with the s orbital and p orbitals to form hybrid bond orbitals. If $N_p$ is the valence of an element, $N_p$ electrons would fill up unpaired in the bond orbitals and would be available for the formation of $N_p$ electron-pair bonds with other atoms. Melting and boiling points of elements would therefore be proportional to $N_p$. Pauling's model could give a qualitative explanation of many properties of transition metals (including those of the Pd and Pt groups), such as interatomic distance, characteristic temperature, hardness, compressibility and coefficient of thermal expansion, and it accounted satisfactorily for the observed values of the atomic saturation magnetic moments of the ferromagnetic metals iron, cobalt, and nickel and their alloys. However, the numerical values that he had derived for the valences of metals were approximate limiting the possibility of making quantitative predictions to validate his model.

Miedema et al. [4-6] developed an equation as below, which predicts the signs of the heat of formation ($\Delta H$) of binary metallic systems,

$$\Delta H = \left[-(\Delta\phi)^2 + \frac{Q}{P}\left(\Delta N^{1/3}\right)^2 - \frac{R}{P}\right]. \qquad \ldots(1)$$

$\phi$ is proportional to Pauling's electronegativity, and $N$ is obtained from the experimental bulk modulus of elements. $\Delta$ denotes the difference in the quantities for two elements. The first term in Eq. (1) is related to the ionicity in the bonds. The parameters $\phi$ and $N$ were adjusted by small amounts to predict the signs of the heats of formation of more than 500 metallic binary systems *with 100% accuracy*. In Eq. (1), the form of its positive term [5], as well as all its constants $\frac{Q}{P}$ and $\frac{R}{P}$, were derived empirically. The success of Eq. (1) confirms the need for a positive term in the equation for $\Delta H$, in addition to the negative term of the form proposed by Pauling. Equation (1) is essentially empirical and there is scope for a reinterpretation of the parameters $\phi$ and $N$ introduced by Miedema et al.

2. **V**alence **and electronegativity of metallic elements**

Miedema et al.'s $N$ are numerically close to the valences proposed by Pauling ($N_P$) [3, 7] for metallic elements. They reproduce the trends in the variation of the melting points and boiling points of elements across the periods in the periodic table *much more closely* than does Pauling's valences, and are thus better suited for use as valence. This fact is illustrated in Figs. 1-3.

Pauling had derived [8] the values of the electronegativity of elements from the energy of their heteronuclear bonds. It had been shown [9] that these values can be obtained from the valence of the elements, if electronegativity is defined as the electrostatic potential due to nuclear charges on the valence electrons. An assumption was made [9] regarding the extent of shielding of nuclear charges by the electrons of an atom for the above purpose.

The bonding in metallic elements is different from that in other elements. The current view is that it is an extremely delocalized communal form of covalent bonding, and we can assume a spherical distribution of valence electrons around the nucleus of an atom. This assumption could be relevant to the way nuclear charges are screened by the electrons of an atom in a metal. We hypothesize in this paper that the screening of nuclear charges by the electrons of an atom in a metal is such that the effective charge felt by a valence electron situated at the midpoint of a bond to the next atom is $N^{1/3}$, where N is the valence of the metal, i.e. all except $N^{1/3}$ electrons exert their full screening power. This means that at any moment in time, all electrons of the atom except $N^{1/3}$ are distributed with spherical symmetry in positions closer to the nucleus than the midpoint of the bond. This can be expected due to the high ligancy and symmetry seen in metals [10], and due to resonance of electrons among various allowed equivalent bond positions. As is done with various hypotheses on screening by electrons in different physical situations [9, 11], this assumption needs to be proven by comparison of its predictions with experimental results. The electrostatic potential at the midpoint of the bond would then be proportional to $N^{1/3}/R$, where R is the metallic radius of the atom. The electronegativity of a metallic atom would then be $\chi_M = Z_{\text{eff}} \, e/R = k \, N^{1/3}/R$, where k is a constant. In Fig. 4(a), we show a plot of Pauling's electronegativity $\chi$ of the metallic elements versus $N_p^{1/3}/R$. A very approximate linear dependence between the two can be observed.

Pauling had divided [12] the metallic elements into three groups: hypoelectronic elements which have fewer electrons than there are orbitals, hyperelectronic ones which have more electrons than there are orbitals and buffer elements. In Fig. 4(b), we show that an almost perfect (R=0.98) linear correlation is obtained for hypoelectronic elements if one plots Miedema's electronegativity values - which have been obtained by correcting Pauling's values to suit the conditions existing in a metal using thermo-chemical data - versus $N^{1/3}/R$, where N are Miedema's *corrected* electron density parameters. The electronegativity values used above have been scaled down to the range of Pauling's electronegativity values using the linear relationship between the two,

$$\chi_M = (\phi - 0.662)/\ 2.157. \qquad \ldots(2).$$

The hypoelectronic elements consist of alkali metals, rare earths and early transition metals. The improvement in linearity for hypoelectronic elements in Fig. 4(b) which uses the *corrected* Miedema parameters is remarkable when compared with Fig. 4(a). In Gordy's treatment of non-metallic elements [9], the elements Cu, Ag and Au were exceptions to the observed correlation. He had noted that Cu, Ag and Au form compounds in which they contribute more than one electron to the formation of covalent bonds: they could be exceptions because the assumption of a shielding constant of unity for all electrons except the ones in a final shell may not be quite appropriate for them. In Fig. 4(b), the buffer elements Ag, Au, Pt, Pd, Rh, Re, Os, and Tc and the hyperelectronic p-metals are found grouped separately from the hypoelectronic elements due to the difference in their electronic structure. The former are characterized by paired electrons in the final shell which do not contribute to valence and could possibly contribute less to shielding and thus increase the electrostatic potential at the midpoint of the bond. The observed linear correlation strongly supports the assignment of $\chi_M$ and $N$ as the more accurate electronegativity and valence respectively of metallic elements. The Miedema electronegativity of metallic elements can be calculated from their Miedema valences as the electrostatic potential at the midpoint of the bond.

3. **Electrical conductivity of transition metals**

With the identification of Miedema's parameters $N$ and $\chi_M$ as valence and electronegativity respectively of metallic elements, the possibility of predicting physical properties of metallic systems within Pauling's model, opens up. Reliable predictions would support Pauling's picture of the electronic structure of metallic elements.

Let $O$ be the total number of orbitals of an atom (see Table I); Pauling had argued that $O = 9$, five 3d, one 4s, and three 4p for the iron group elements, and corresponding sets for other series [3]. Let $E$ be the total number of electrons beyond the inert gas shell of an atom. In the case of atoms with less than 5 electrons beyond the inert gas shell, $O = 5$, i.e. there is no admixture of the s and p orbitals with the d orbitals. Let $N$ be the valence of a metal; Pauling had proposed that $N$ orbitals would be filled by an equal number of unpaired electrons which would form resonating covalent bonds with the neighboring atoms. The number of electrons not involved in bonding would be $E_P = E - N$. They (the $E_P$ electrons) would fill remaining orbitals in pairs. Such electrons were referred to as paired non-bonding electrons by Pauling. In the case of atoms of Fe, Co and Ni with magnetic moments $M$ $\mu_B$ per atom [3], M electrons would also fill unpaired and would be localized on the atom without taking part in the bonding, and the rest $(E - N - M)$ would fill paired. Let the number of orbitals occupied by the paired (or (paired + M), in the case Fe, Co and Ni) electrons be $O_P$; $O_P = (E - N - M)/2$. Pauling had recognized the need for empty orbitals to facilitate metallic conduction [13] and he called them 'metallic orbitals'. He assumed 0.72 metallic orbitals uniformly for all elements. Instead one can determine the number of metallic orbitals as $O_M = O - (N + O_P)$. The electrical conductivity of an element ($\sigma$) would be proportional to the product of the number of electrons available after bonding and the number of metallic orbitals available for them during the conduction process, i.

e. to $E_P$ x $O_M$. The values of $E_P$ x $O_M$ are listed in the Table I. The experimental values of electrical conductivity of transition metals $\sigma_{ex}$, from Kittel's book [14], are listed in the last column of the Table.

The experimentally obtained electrical conductivity of transition metals is plotted in Fig. 5, against $E_P$ x $O_M$ from Table I. We observe from the figure that the electrical conductivity of transition metals increases linearly with $E_P$ x $O_M$. It is to be noted that the present approach predicts, for the first time, that the electrical conductivity of transition metals would decrease in the order $\sigma_{Ag} > \sigma_{Cu} > \sigma_{Au} > \sigma_{Rh} > \sigma_{others}$, as is experimentally observed. The elements W, Mo, Mn, and Pd (out of a total of 27 transition metals) do not fit the trend may be due to possible intricacies in their band structure which make them exceptions in the simple picture presented herein.

4. **Summary and discussion**

We have in this paper pointed out *the possibility* that Miedema's parameters $\phi$ and $N$ can be identified as the electronegativity and valence of metallic elements. Rigorous interpretation of this conclusion using ab initio theoretical methods is of interest. Efforts to understand such lumped or average constants of elements using Density Functional Theory have been reported in the literature [15, 16].

The present work suggests that Pauling's scheme of describing the electronic structure of metallic elements could essentially be correct, if the metallic valences can be considered as given by Miedema's parameter $N$. Once the valences of metals are defined with good accuracy, the number of metallic orbitals need not be assumed to be a constant at 0.72 for all metals as Pauling had done. Pauling's model then reproduces the trends in the variation of electrical conductivity of transition metals with good accuracy.

One might recall that the parameters of Miedema et al. can be used to predict the signs of the heats of formation of metallic alloys with excellent accuracy [4]. We have shown elsewhere [17] that the positive term in the heat of formation Eq. (1) stems from the energy expended in charge transfer from the more electronegative atom to the leas electronegative atom to maintain electroneutrality. With the hypothesis that the parameters $\phi$ and $N$ are respectively the electronegativity and valence of metallic elements, the extension of Pauling's model could predict the sign of the heat of formation of metallic alloys with excellent accuracy, with the value of Miedema's empirical constant Q/P following from the model [17]. Rajasekharan and Girgis had shown that the binary systems with intermetallic compounds of the same crystal structure fall on a straight line on a ($\Delta\phi$, $\Delta N^{\frac{1}{3}}$) map [18, 19]. We have recently shown by an examination of the 96 crystal structure types in which ~ 3000 intermetallic compounds crystallize that the map using the parameters $\phi$ and $N$ can be used to identify concomitant and mutually exclusive structure types in binary phase diagrams with high accuracy [20, 21]. Since the crystal structures of intermetallic compounds are likely to be decided by a number of competing energy contributions, our finding that structure predictions are possible using just two parameters per element is indeed surprising. An interpretation of that observation has been attempted in [20]. Knowledge of the quantities $\phi$ and $N$ *to their quoted accuracy*, up to the second decimal place, is necessary make the above predictions as well as to make predictions regarding the electrical conductivity of transition elements as done in the present paper. The above observations point to the importance of the quantities, $\phi$ and $N$ in deciding the properties of metallic systems. Since properties of the solid state are decided by the electronic structure [2], the relevance of these parameters to electronic band structure of metallic solids becomes evident.

The conclusion from the ability of Pauling's model to predict the outcome of electrical conductivity experiments would be that cohesion in metallic elements is due to a certain number ($N$) of electrons occupying an equal number of orbitals, and forming bonds between atoms; and that the bands constituted by the bonding orbitals are full and do not contribute to electronic properties. The electronic properties are decided by the partially filled band formed by the rest of the orbitals. The contribution of carrier density to electrical conductivity seems to dominate differences due to scattering. It is interesting to note that the numerical value of $N$ is transferrable from the elements to their alloys. Band structures of metallic elements and their compounds are now calculated using ab initio techniques. There are applications of Pauling's RVB theory using ab initio methods [22]. It is of interest to examine whether Pauling's' description of the electronic structure of elements and alloys [3], with Miedema's $N$ as the valence, can be validated by such calculations.

5.  **Acknowledgements**

TR thanks DMRL, Hyderabad, for permission to publish this paper. VS thanks CSIR for research funding (03(1065) 106/EMR-II)

**Table I** The electronic structure of transition elements in the metallic environment, and their electrical conductivity values [13] $\sigma_{ex}$ in mho.m$^{-1}$.

|     | O | E  | N     | $E_P = E - N$ | M   | $O_P$ | $N + O_P$ | $O_M$ | $E_P \times O_M$ | $\sigma_{ex}$ |
|-----|---|----|-------|---------------|-----|-------|-----------|-------|------------------|---------------|
| Sc  | 5 | 3  | 2.048 | 0.952         | 0   | **0.476** | 2.524 | 2.476 | 2.357            | 0.21          |
| Ti  | 5 | 4  | 3.512 | 0.488         | 0   | 0.244 | 3.756     | 1.244 | 0.607            | 0.23          |
| V   | 5 | 5  | 4.411 | 0.589         | 0   | 0.295 | 4.706     | 0.295 | 0.173            | 0.5           |
| Cr  | 9 | 6  | 5.178 | 0.822         | 0   | 0.411 | 5.589     | 3.41  | 2.804            | 0.78          |
| Mn  | 9 | 7  | 4.177 | 2.823         | 0   | 1.412 | 5.589     | 3.412 | 9.631            | 0.072         |
| Fe  | 9 | 8  | 5.545 | 2.455         | 2.2 | 2.328 | 7.873     | 1.128 | 2.768            | 1.02          |
| Co  | 9 | 9  | 5.359 | 3.641         | 1.6 | 2.621 | 7.980     | 1.021 | 3.716            | 1.72          |
| Ni  | 9 | 10 | 5.359 | 4.641         | 0.6 | 2.621 | 7.980     | 1.021 | 4.736            | 1.43          |
| Cu  | 9 | 11 | 3.177 | 7.823         | 0   | 3.912 | 7.089     | 1.912 | 14.954           | 5.88          |
| **Y** | 5 | 3 | 1.772 | 1.228         | 0   | 0.614 | 2.386     | 2.614 | 3.210            | 0.17          |
| Zr  | 5 | 4  | 2.686 | 1.314         | 0   | 0.657 | 3.343     | 1.657 | 2.177            | 0.24          |
| Nb  | 5 | 5  | 4.252 | 0.748         | 0   | 0.374 | 4.626     | 0.374 | 0.280            | 0.69          |
| Mo  | 9 | 6  | 5.545 | 0.455         | 0   | 0.228 | 5.773     | 3.228 | 1.469            | 1.89          |
| Tc  | 9 | 7  | 5.93  | 1.07          | 0   | 0.535 | 6.465     | 2.535 | 2.712            | 0.7           |
| Ru  | 9 | 8  | 6.128 | 1.872         | 0   | 0.936 | 7.064     | 1.936 | 3.624            | 1.35          |
| Rh  | 9 | 9  | 5.452 | 3.548         | 0   | 1.774 | 7.226     | 1.774 | 6.294            | 2.08          |
| Pd  | 9 | 10 | 4.657 | 5.343         | 0   | 2.672 | 7.329     | 1.672 | 8.931            | 0.95          |
| Ag  | 9 | 11 | 2.687 | 8.313         | 0   | 4.157 | 6.844     | 2.157 | 17.927           | 6.21          |
| La  | 5 | 3  | 1.295 | 1.705         | 0   | 0.853 | 2.148     | 2.853 | 4.864            | 0.13          |
| Hf  | 5 | 4  | 2.924 | 1.076         | 0   | 0.538 | 3.462     | 1.538 | 1.655            | 0.33          |
| Ta  | 5 | 5  | 4.331 | 0.669         | 0   | 0.335 | 4.666     | 0.335 | 0.224            | 0.76          |
| W   | 9 | 6  | 5.93  | 0.07          | 0   | 0.035 | 5.965     | 3.035 | 0.212            | 1.89          |
| Re  | 9 | 7  | 6.435 | 0.565         | 0   | 0.283 | 6.718     | 2.283 | 1.290            | 0.54          |
| Os  | 9 | 8  | 6.332 | 1.668         | 0   | 0.834 | 7.166     | 1.834 | 3.059            | 1.1           |
| Ir  | 9 | 9  | 6.128 | 2.872         | 0   | 1.436 | 7.564     | 1.436 | 4.124            | 1.96          |
| Pt  | 9 | 10 | 5.64  | 4.36          | 0   | 2.18  | 7.82      | 1.18  | 5.145            | 0.96          |
| Au  | 9 | 11 | 3.87  | 7.13          | 0   | 3.565 | 7.435     | 1.565 | 11.158           | 4.55          |

The symbols used as headings of all the columns are explained in the text.

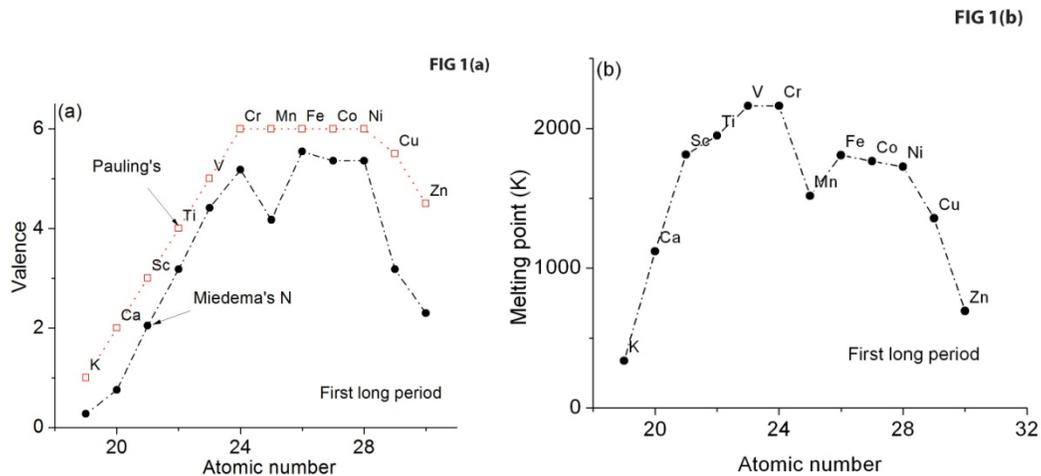

**Fig. 1.** Comparison of Miedema's *N* with Pauling's valences ($N_p$) for metals; **a)** We see that, in the first long period of the periodic table, Miedema's *N* are comparable in magnitude, and vary in the same way with the atomic number of the elements, as do Pauling's valences. In **b),** we see that *N* follows the variation in the melting points of elements, including the anomalous behaviour of Mn, more closely than Pauling's valences. The numerical values of Pauling's valences can be adjusted by a constant value to match *N* even better.

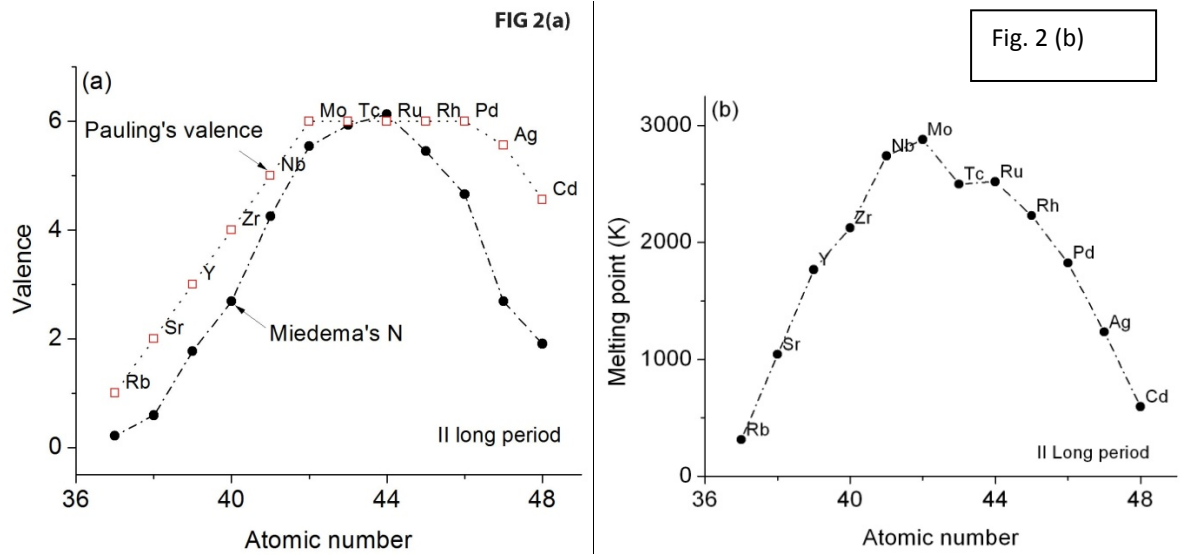

**Fig. 2.** Comparison of Miedema's *N* with Pauling's valences ($N_p$) for metals; **a)** We see that, in the second long period of the periodic table, Miedema's *N* are comparable in magnitude, and vary in the same way with the atomic number of the elements, as do Pauling's valences. In **b)**, we see that *N* follows the variation in the melting points of elements more closely than Pauling's valences.

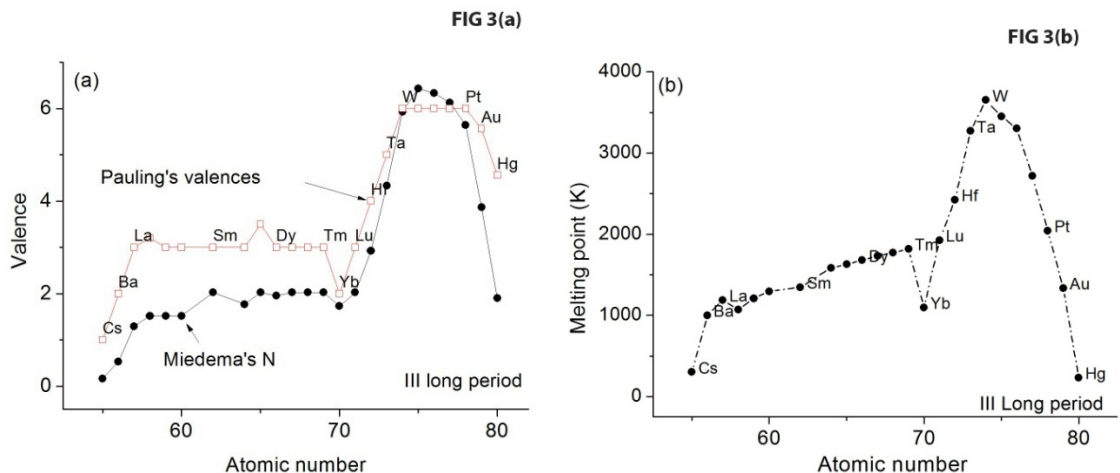

**Fig. 3.** Comparison of Miedema's *N* with Pauling's valences (N$_p$) for metals; **a)** We see that, in the first long period of the periodic table, Miedema's *N* are comparable in magnitude, and vary in the same way with the atomic number of the elements, as do Pauling's valences. In **b)** we see that *N* follows the variation in the melting points of elements more closely than Pauling's valences.

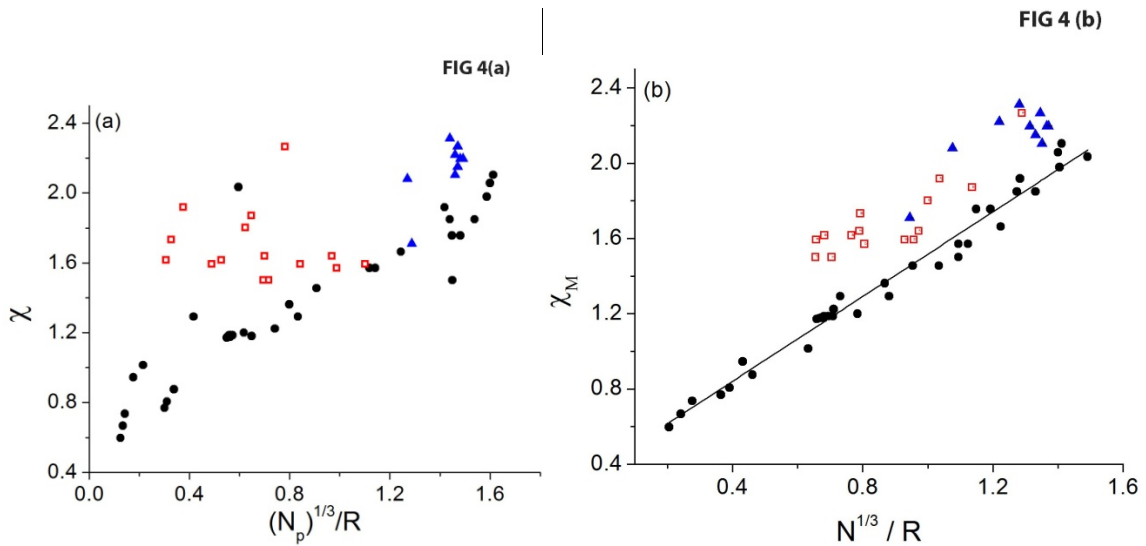

**Fig. 4.** Effect of corrections using the experimental information on the signs of heats of formation on Miedema's $N$ and $\phi$; **a)** We plot Pauling's $\chi$ versus $N_P/R$, where $N_P$ are Pauling's metallic valences and R is the metallic radius. In **b)**, a plot of $\chi_M$ versus $(N^{1/3})/R$ is shown; $\chi_M$ are Miedema's electronegativity values (obtained from $\phi$), and $N$ are the electron density parameter of Miedema. The plot in (**b**) has absorbed the effect of the corrections made by Miedema et al. and the effects are dramatic and clearly not accidental. A good linearity with a regression factor of 0.98 is observed for the hypoelectronic transition metals, s-elements and rare earths (closed circles, black online) supporting the hypothesis that $N^{1/3}$ can be considered the valence of the elements effective on a linear bond, and $\chi_M$ the electronegativity. The p-metals which are hyperelectronic (seen as unfilled squares, red online) and the buffer elements Ag, Au , Re, Ru, Os, Tc, Rh, Ir, Pt, and Pd (triangles, blue online) form separate groups.

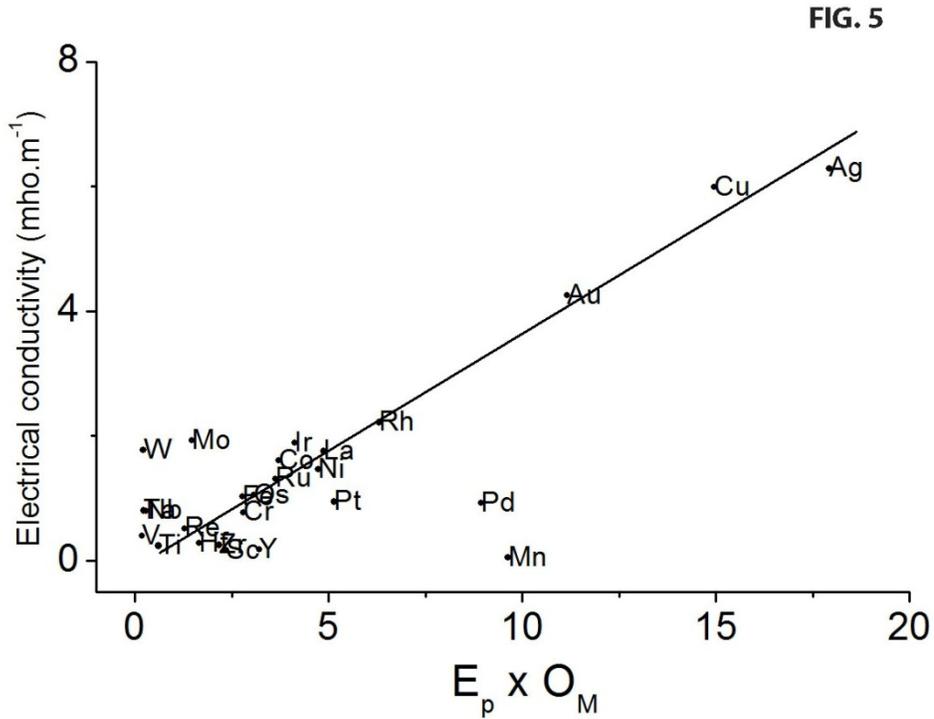

**Fig. 5.** Electrical conductivity of transition metals is plotted versus $E_P \times O_M$. A linear dependence is seen, with the conductivity decreasing in the order $\sigma_{Ag} > \sigma_{Cu} > \sigma_{Au} > \sigma_{Rh} > \sigma_{others}$. $N$ electrons fill the bonding orbitals unpaired, where $N$ is Miedema's electron density parameter. In the case of Fe, Co and Ni, the magnetic electrons are also localized on the atoms and fill a few more orbitals unpaired. $E_P$ is the number of remaining electrons, the paired non-bonding electrons of Pauling. $O_M$ is the number of unfilled orbitals. The values of $E_P \times O_M$ are listed in Table I. The experimental values of electrical conductivity of transition metals $\sigma_{ex}$ are from Kittel's book [14].

**Figure captions**

**Fig. 1.** Comparison of Miedema's *N* with Pauling's valences ($N_p$) for metals; **a)** We see that, in the first long period of the periodic table, Miedema's *N* are comparable in magnitude, and vary in the same way with the atomic number of the elements, as do Pauling's valences. In **b)**, we see that *N* follows the variation in the melting points of elements, including the anomalous behaviour of Mn, more closely than Pauling's valences. The numerical values of Pauling's valences can be adjusted by a constant value to match *N* even better.

**Fig. 2.** Comparison of Miedema's *N* with Pauling's valences ($N_p$) for metals; **a)** We see that, in the second long period of the periodic table, Miedema's *N* are comparable in magnitude, and vary in the same way with the atomic number of the elements, as do Pauling's valences. In **b)**, we see that *N* follows the variation in the melting points of elements more closely than Pauling's valences.

**Fig. 3.** Comparison of Miedema's *N* with Pauling's valences ($N_p$) for metals; **a)** We see that, in the first long period of the periodic table, Miedema's *N* are comparable in magnitude, and vary in the same way with the atomic number of the elements, as do Pauling's valences. In **b)** we see that *N* follows the variation in the melting points of elements more closely than Pauling's valences.

**Fig. 4.** Effect of corrections using the experimental information on the signs of heats of formation on Miedema's *N* and $\phi$; **a)** We plot Pauling's $\chi$ versus $N_P/R$, where $N_P$ are Pauling's metallic valences and R is the metallic radius. In **b)**, a plot of $\chi_M$ versus $(N^{1/3})/R$ is shown; $\chi_M$ are Miedema's electronegativity values (obtained from $\phi$), and *N* are the electron density parameter of Miedema. The plot in (**b**) has absorbed the effect of the corrections made by Miedema et al. and the effects are dramatic and clearly not accidental. A good linearity with a regression factor of 0.98 is observed for the hypoelectronic transition metals, s-elements and rare earths (closed circles, black online) supporting the hypothesis that $N^{1/3}$ can be considered the valence of the elements effective on a linear bond, and $\chi_M$ the electronegativity. The p-metals which are hyperelectronic (seen as unfilled squares, red online) and the buffer elements Ag, Au , Re, Ru, Os, Tc, Rh, Ir, Pt, and Pd (triangles, blue online) form separate groups.

**Fig. 5.** Electrical conductivity of transition metals is plotted versus $E_P$ x $O_M$. A linear dependence is seen with the conductivity decreasing in the order σ$_{Ag}$ > σ$_{Cu}$ > σ$_{Au}$ > σ$_{Rh}$ > σ$_{others}$. The symbols $E_P$ and $O_M$ are explained in the text.